\documentclass{article}
\usepackage{fortschritte}
\def\beq{\begin{equation}}                     %
\def\eeq{\end{equation}}                       %
\def\bea{\begin{eqnarray}}                     
\def\eea{\end{eqnarray}}                       
\begin {document}
\def\titleline{
%
%
%
Transplanckian Dispersion Relation \\
and\\
Entanglement Entropy of Blackhole
}
\def\email_speaker{
{\tt
cschu@phys.nthu.edu.tw
}}
\def\authors{
%
%
%
%
%
Darwin Chang\1ad ,
Chong-Sun Chu \1ad\sp,                           
Feng-Li Lin \2ad,
}
\def\addresses{
\1ad                                        
Department of Physics, National Tsing-Hua University,
Hsin-Chu, 300 Taiwan  \\
\2ad
Physics Division,
National Center for Theoretical Sciences,\\
National Tsing-Hua University, Hsin-Chu, 300 Taiwan
}
\def\abstracttext{
The quantum correction to the entanglement entropy of the event
horizon is plagued by the UV divergence due to the infinitely
blue-shifted near horizon modes. The resolution of this 
UV divergence provides an excellent window to a better understanding 
and control of the quantum gravity effects. We claim that the key to 
resolve this UV puzzle is the transplanckian dispersion relation.
We calculate the entanglement entropy using 
a very general type of transplanckian dispersion relation such that
high energy modes above a certain scale are cutoff, and show that 
the entropy is rendered UV finite.
We argue that modified dispersion relation is a 
generic feature of string theory, and this boundedness nature of the 
dispersion relation is a general consequence of 
the existence of a minimal distance in string theory. } 

\large \makefront
\section{Introduction and Conclusion}

   The biggest puzzle in the black hole physics is the
origin of the Bekenstein-Hawking entropy \cite{BH} and the related
information loss problem. There have been various proposals to try
to count the microscopic degrees of freedom associated with the
Bekenstein-Hawking entropy, in all cases peoples believe that the
near horizon fluctuations should play an important role. The most
radical one is to single out only the horizon itself by neglecting
the other region of the spacetime and then attribute the
Bekenstein-Hawking entropy to
some kind of dynamics associated with the horizon
hypersurface. For example, the stretched horizon in the membrane
paradigm \cite{r1}, the isolated horizon in the context of loop quantum
gravity \cite{r2} or the conformal field theory associated with the near
horizon symmetry algebra \cite{r3}.

  The other common proposal once adopted is that the
Bekenstein-Hawking entropy arises from the entanglement of the
quantum fluctuations inside and outside the horizon. The reason
for this proposal is that the inside region of the horizon is
causally disconnected from the outside one, then for the outside
observer it is natural to trace over the Hilbert space of the
inside region and obtain an entangled state. The entropy
associated with these entangled states is called the entanglement
entropy of the black hole.

There are various techniques developed to calculate the
entanglement entropy of the black hole. One is the
brick wall method introduced by 't Hooft \cite{hooft}. This
method treats the horizon as a brick wall such that the
wavefunction strictly vanishes inside the horizon, this is
equivalent to the fact that no information associated with the
quantum fluctuation inside the horizon can be carried out to the
outside observer, and therefore, all the quantum states obeying
this boundary condition are entangled. We should emphasize that
the above equivalence is only in a heuristic way and a rigorous
proof is absent. However, we will take the brick wall method as
the starting point in calculating the entanglement entropy without
questioning the above underlying assumption of the equivalence.

 If one thinks that the fluctuations of the entangled states are
the quantized gravitational fluctuations on the background
spacetime, then the entanglement entropy could be the origin of
the Bekenstein-Hawking entropy. However, this expectation turns
out to be too naive.  In quantum field theory in flat or curved
space, the quantum effect is usually plagued by the UV divergence
which is taken care by proper renormalization scheme, however,
once gravity is included, the UV divergence will cause enormous
back reaction of the background geometry and cannot be justified
by the renormalization scheme. The famous example is the
well-known cosmological constant problem. Similar problem happens
in calculating the entanglement entropy which turns out to be UV
divergent so that an infinitesimal brick wall thickness is
introduced as an UV cutoff. The brick wall cutoff indicates that
the near horizon fluctuations usually get infinitely blue-shifted,
which will then cause a large back reaction to themselves and to
the background. Once the effect of the back reaction is taken into
account, the area law may or may not hold true. If the area law is
violated we hope that the entanglement entropy contributes only a
small correction to the Bekenstein-Hawking entropy whose area law
nature is suggested by the holographic principle, and may be
related to some unknown nature of quantum gravity.

This puzzle of UV divergence should be taken seriously because it
is a reflection of our lack of understanding and control 
of quantum gravity effects.
From a bottom-up approach, one
may be able  to get better understanding into the nature of quantum gravity 
from the possible low energy 
effective features that can resolve the UV
divergence problem of the event horizon.  Such feature may point
its finger toward the essential aspects of quantum gravity.  In this
paper, we propose that transplanckian dispersion relations(TDRs)
is one such feature.

In our paper \cite{CCL}, we modeled the effect of the back reaction
mentioned earlier by modifying the propagator of the fluctuations by the
so called transplanckian dispersion relations(TDRs), and then calculate
the transplanckian entanglement entropy based on the modified dispersion
relations. In principle, one should be able to determine how quantum
gravity effects would modify the dispersion relation from string theory.
Unfortunately, our present technology in string theory is not powerful
enough to allow one to quantize string theory in a blackhole background.
Nevertheless, it is extremely natural to expect for TDRs in string
theory. For example, in the simpler case of open string in a constant
$B$-field background, a modified dispersion relation is resulted due to
the IR/UV mixing effect\cite{sei}. The resulting dispersion relation
respects the symmetry of the background, which is smaller than the full
Lorentz symmetry of the flat spacetime. It is clear that in the case of
closed string, the dispersion relation will also generally get modified
due to string loop effects. See, for example, \cite{bb} the string
theory calculation in the case of open string in $B$-field.

Although it is still technically impossible at present to derive the TDRs
from first principle string theory, it can be argued that
\cite{fawad,mersini} due to the existence of a minimal length scale
$l_s$ in string theory, the TDRs is expected to cutoff the high energy
mode with energy above $k_0 \sim 1/l_s$.
We will thus consider a generic class of the TDRs of such kind instead of
some specific one. Our conclusion is that for such a class of TDRs which
are bounded by some reasonable energy scale, the entanglement entropy is
UV finite. Thus in particular the entanglement entropy is always finite 
in string theory.
We also found that the area law nature is lost with 
its value negligible in
comparison with the Bekenstein-Hawking entropy in the semi-classical
limit.

\section{Transplanckian Dispersion Relation}

The dispersion relation of a particle is a relation between its
energy and momentum. At the tree level, it can be obtained from
the invariant of the Lorentz group, and for the following
background metric considered in this paper
\begin{equation} \label{metric} ds^2 = g_{00} dt^2 + g_{rr} dr^2 +
f d\Omega_2^2,
\end{equation}
it takes the following form
\begin{equation} \label{disperse1}
g^{00}\omega^2+g^{rr}p^2_r+p_{\perp}^2=0\;,
\end{equation}
where $g^{00} =1/g_{00}, g^{rr}= 1/g_{rr}$ and
$p_{\perp}^2=g^{mn}p_mp_n = \ell(\ell +1)/f$ is the transverse
momentum squared and $\ell$ is the angular momentum quantum
number. Note that $g_{00}, g_{rr}$ and $f$ are functions of $r$.

It is convenient to introduce the following definitions
\beq
\label{new-def} \xi^2 := p_\perp^2 g_{rr}= {\ell(\ell
+1)g_{rr}\over f}, \quad \rho^2 := - {g_{00} \over g_{rr}},
\eeq
the dispersion relation (\ref{disperse1}) takes the form
\beq\label{p-H0}
p_r=\sqrt{{\omega^2 / \rho^2}-\xi^2}.
\eeq

  As mentioned in the Introduction, the large back reactions due
to the infinitely blue-shifted near horizon modes are supposed to
be important in evaluating the entanglement entropy. Moreover, it
is quite generally believed that nonlocal effects due to quantum
gravity will provide a natural regulator to the UV divergence in
quantum field theory by suppressing the contributions of the high
energy modes. The simplest proposal to encode the effects of this
suppression is to replace the linear dispersion relation by the
so-called transplanckian dispersion relation(TDR)
\cite{Unruh,Corley}.

Consider a spherically symmetric background given by the general
metric (\ref{metric}), it is reasonable to impose the TDR
according to the residual spacetime symmetry preserved by the
metric.  Since the transplanckian effect will be mostly due to the
blue-shift in the near horizon regime in the radial direction, so
it is natural to impose TDR along that direction to suppress the
blue-shift effect. Therefore we consider TDR of the form
\beq \label{disperse2}
g^{00}\omega^2+g^{rr}H^2(p_r)+p_{\perp}^2=0\;.
\eeq
The function $H$ describes the transplanckian effects. In terms of
the $\rho$ and
$\xi$ variables, the TDR (\ref{disperse2}) then takes the form
\begin{equation} \label{p-H}
p_r= H^{-1} (\sqrt{{\omega^2/ \rho^2}-\xi^2} )\;.
\end{equation}

For example,  Unruh \cite{Unruh} and respectively, Corley and
Jacobson (C-J) \cite{Corley} proposed respectively:
\beq 
 H_n(k)=k_0[\tanh({k / k_0})^n]^{1\over n}; \qquad
H(k)=\sqrt{k^2-{k^4/( 4k_0^2)}}, \quad k \le 2
k_0\;. \label{disp-corley}
\eeq
The corresponding radial momentum is
\beq
p_r =k_0 \{\tanh^{-1}[k_0^{-2}({\omega^2/ \rho^2}
-\xi^2)]^{n\over 2}\}^{1\over n}; \qquad
p^{\pm}_r=\sqrt{2} k_0\sqrt{1\pm \sqrt{1-k_0^{-2}
({\omega^2/ \rho^2}-\xi^2)}}. 
\eeq
Note that the Unruh's proposal does not cutoff the high momentum
modes in contrast to the C-J's case, but just suppresses the high
energy modes.

Despite that both Unruh's and C-J's proposals have been used
generally in the different contexts, it is not quite possible that
the TDRs from the higher theory will take those specific forms. 
As discussed above, string theory suggests generally that the TDRs 
satisfy  the following boundedness condition:
\beq\label{bound}
0\le H^2(k) \le k_0^2,
\eeq
for some $k_0$, and it is natural from string theory that 
$k_0 \sim 1/l_s$. We will call these {\it bounded-TDRs}. 
It follows that for this bounded class of TDR
\beq \label{energy} \max({\omega^2/ \rho^2}-k_0^2,0) \le \xi^2 \le
{\omega^2 /\rho^2}\;.
\eeq
The suppression of the high energy modes is 
effected by the bounded function $H$, whose explicit form  will depend 
on the details of the quantum gravity effects. In \cite{CCL} we showed
that without knowing explicitly the form of $H$, the transplanckian 
entanglement entropy is always rendered UV finite. Thus we conclude that the 
entanglement entropy is UV finite for the bounded TDRS, in particular 
in  string theory.

\section{Entanglement Entropy of Black Hole}
  Using the  dispersion relation (\ref{disperse1}), the
entanglement entropy of the black hole was found to be divergent
due to the infinitely blue-shifted near horizon region
\cite{hooft}. In the brick wall model, a UV cutoff is introduced
to regularize the divergence by imposing the condition on the
wavefunction $\Phi$ so that 
$ \Phi(r)=0$, for $r \le r_h+\epsilon$,
where $\epsilon$ is the infinitesimal brick wall thickness
\cite{hooft}. This has the effect of {\it cutting out} the near horizon
modes.

The total free energy is obtained by summing over the
contributions from all the physical modes satisfying the brick
wall condition,
\beq \label{free1}
\beta F = \int d\omega \; z(\beta \omega) {dg(\omega) \over d\omega}
\eeq
where the Boltzmann weight $z(x):=\ln(1-e^{-x})$, $\beta$ is the
inverse temperature, and $g(\omega)$ is the density of states. In
the WKB approximation \cite{hooft},  it is
\beq\label{density}
\pi g(\omega) = \int d\rho \mu(\rho) d\xi \;  \xi \; p_r,
\eeq
where $ \mu(\rho) := {2 f \over g_{rr}}{dr\over d\rho}$ is a ``measure
factor''.
For fixed $\omega$ and $\rho$, we require $\xi \le \omega/\rho$ to
have $p_{r}$ real. This condition should be imposed in the
$\xi$-integration.

 Consider the black hole background
\beq
\label{bh1} ds^2=-h(r)dt^2+h(r)^{-1}dr^2+r^2 d\Omega_2^2,
\eeq
where $0< h(r)=1-{ 2M /r} < 1$.
In this case $\rho=h$ and
$\mu(\rho)=2(2M)^3 \rho/(1-\rho)^4$.
The density of state is
\beq
g(\omega) = {2 (2M)^3\omega^3 \over 3 \pi} ({1\over \epsilon} + {1\over 3
\delta^3}+\cdots) ,  \label{g-bh}
\eeq
where we have introduced an
UV (resp. IR) cutoff $\epsilon$ (resp. $\delta$) in the
$\rho$-coordinate, which corresponds to an UV (resp. IR) cutoff
$r=2M/(1-\epsilon)$ (resp. $r= 2M/\delta$ ) in the $r$-coordinate;
and $\cdots$ are the sub-leading terms in the limit of $\delta,
\epsilon \rightarrow 0$. Using (\ref{g-bh}) and $\beta = 8 \pi M$,
we get
\beq \label{hooft-bh} F={-2\pi^3 (2M)^3 \over 45 \beta^4}({1
\over \epsilon} + {1\over 3\delta^3}), \quad
S={1\over 360}({1 \over \epsilon} + {1\over 3\delta^3})
\eeq
which is plagued by both UV and IR divergences. This agrees with
the eqn. (3.12) of \cite{hooft} except that a IR divergent
boundary contribution had been dropped there. The IR-divergent
piece is independent of $M$ and represents the contribution from
the vacuum. In \cite{hooft}, the UV-divergent piece was shown to
give an entropy which obeys the area law
\beq \label{Suv}
S = {A\over 360\pi \epsilon^2_{\rm p}}
\eeq
if the UV-cutoff is given in terms of the proper distance, i.e.
$\epsilon_{\rm p}\approx 4M\sqrt{\epsilon}$. In modern language,
the relation between $\epsilon$ and $M$ for fixed $\epsilon_{\rm
p}$ required by the area law indicates a holographic nature of
UV/IR connection \cite{UVIR}.

 We should mention that the UV divergent area law formula (\ref{Suv})
is universal, namely, the form is the same for the other event
horizons such as the ones in the Rindler space \cite{Susskind} and
in the de Sitter space \cite{CCL}. This indicates the near horizon
dynamics dictating the entanglement entropy is universal.

\section{Transplanckian Black Hole Entropy}
Our goal is to determine the UV behavior of $F$ and $S$ of the
Schwarzschild black hole for a generic bounded TDR with a $H$
satisfying (\ref{bound}). Combining the condition (\ref{energy})
with the brick wall condition
$\rho\ge \epsilon$, we have
the following
integration branches contributing to the free energy (up to a
factor of $1/(\beta \pi)$)
\bea
&(1)& \int_{\epsilon k_0}^{(1-\delta)k_0} d\omega  \;z
\frac{d}{d\omega} \int_\epsilon^{\omega / k_0} d\rho \mu(\rho)
\int_{\sqrt{{\omega^2/\rho^2}-k_0^2}}^{\omega / \rho} d\xi \xi
p_r,
\;\;  \label{br1} \\
&(1*)& \int_{(1-\delta) k_0}^{\infty} d\omega \; z
\frac{d}{d\omega} \int_\epsilon^{1-\delta} d \rho  \mu(\rho)
\int_{\sqrt{{\omega^2 / \rho^2}-k_0^2}}^{\omega / \rho} d\xi \xi
p_r ,
\;\;\;\;\;\; \label{br1s}\\
&(2)&  \int_{\epsilon k_0}^{(1-\delta)k_0} d\omega \;z
\frac{d}{d\omega} \int_{\omega / k_0}^{1-\delta} d\rho\mu(\rho)
\int_0^{\omega/\rho} d\xi\xi p_r , \label{br2}\\
&(3)& \int_0^{\epsilon k_0} d\omega \;z  \frac{d}{d\omega}
\int_{\epsilon}^{1-\delta}  d\rho \mu(\rho) \int_0^{\omega/\rho}
d\xi \xi p_r.
\label{br3}
\eea
Note that by comparing the ranges of the above $\xi$-integrations
with the ones without transplanckian suppression, we find that the
use of TDR leads to a reduction of the allowed angular momentum
modes. This leads to a suppression on the density of states and
hence
eventually to the UV finiteness of the entropy as shown below.

It is convenient to change variable
$x=\sqrt{\omega^2/\rho^2-\xi^2}$ in order to isolate the
$\omega$-dependence in $g(\omega)$. We have
\beq
\label{br}
\pi g(\omega)=  \int_{\cdot}^{\cdot} d\rho\;
\mu(\rho)\int_{\cdot}^{\cdot} dx \;x H^{-1}(x)\;.
\eeq
Note that the integrand does not depend on
$\omega$ and $\epsilon$,
but the integration limits may.  Note also that the $(1*)$-branch
only exists for BH and de Sitter due to the additional constraint
$h\le 1$.
One may try to estimate the UV behavior for the integrals for each
branches. However this is rather complicated technically.
Surprisingly, without knowing explicitly these integrals, one can
show that the sum of the branches (1), (1*), (2) and (3), i.e. the
free energy $F$, is completely UV finite!

To demonstrate this, let us first consider (1*) branch. The
integration limits are independent of $\omega$, i.e.
$\int^{1-\delta}_{\epsilon} d\rho \int^{k_0}_0 dx$, so is $g(\omega)$.
Therefore (1*) does not contribute. For (1) branch the
$dx$-integral is independent of $\omega$, however, there is
$\omega$ dependence for $g(\omega)$ coming from the integration
limits for $d \rho$-integral. Using the fundamental theorem of
calculus, we can carry out the derivative w.r.t $\omega$ in
(\ref{br1}) and obtain
\beq \label{F1}
F_{(1)}={1\over \pi \beta k_0}\int^{k_0}_0 dx x H^{-1}(x) \cdot
\int^{(1-\delta)k_0}_{\epsilon k_0} d\omega
\;z(\beta\omega)\;\mu({\omega\over k_0}).
\eeq
Similarly, we can carry out the derivative w.r.t. $\omega$ for
branch (2) and (3), and obtain
\beq
\label{F2}
F_{(2)}=-F_{(1)}+{1 \over \pi
\beta}\int^{(1-\delta)k_0}_{\epsilon k_0} d\omega z
\int^{k_0}_{\omega \over 1-\delta} dx\; \mu({\omega\over x})
H^{-1}(x)\;,
\eeq
\beq
\label{F3}
F_{(3)}={1\over \pi \beta}\int_0^{\epsilon k_0} d\omega z
\int_{k_0}^{\omega\over \epsilon}dx\; \mu({\omega\over x})H^{-1}(x)
+{1\over \pi \beta}\int_0^{\epsilon k_0} d\omega z
\int^{k_0}_{\omega\over 1-\delta} dx\; \mu({\omega\over
x})H^{-1}(x)\;.
\eeq
Physically it is required that the integral
\beq \label{H-cond}
I:=\int^{k_0}_0 dx\; x H^{-1}(x) < \infty
\eeq
to be finite. The reason is because the density of states for
branch (1), in (\ref{br1}), is given by  $\pi g(\omega) = I \cdot
\int_\epsilon^{\omega / k_0} d\rho \mu(\rho) $, and it should be
finite for any sensible physical system. This condition also
guarantees our manipulation above for (\ref{F1}), (\ref{F2}) to
make sense. Of course (\ref{H-cond}) is true for both  Unruh's and
C-J's TDRs.

Next we examine the first term of (\ref{F3}), we perform a change
of variables by $\omega=\epsilon k_0 y$ and obtain
\beq \label{small}
{k_0 \epsilon \over \pi
\beta}\int^1_0 dy \; z(\beta\epsilon k_0 y) \int^{k_0 y}_{k_0} dx \;
\mu({\epsilon k_0 y\over x}) H^{-1}(x) \;.
\eeq
Using the fact $z(x)\approx \ln(x)-x/2+\cdots$ around $x=0$, and
assuming the following condition on the near horizon geometry
\beq\label{meas-cond}
\epsilon \mu(\epsilon)\approx \epsilon^\alpha \quad\mbox{for}\quad
\alpha>0, \quad \mbox{as}\quad \epsilon \to 0,
\eeq
which is true for Schwarzschild, Rindler and de Sitter metrics, we
can manipulate (\ref{small}) by interchanging the order of
integrations over $dx$ and $dy$ so that the $y$-integration can be
carried out, (\ref{small}) then
becomes
$ (\pi \alpha k_0)^{-1}\epsilon^\alpha \ln \epsilon  \int^{k_0}_0
dx\; x H^{-1}(x).$ This vanishes as $\epsilon \to 0$.

Finally, the second term of (\ref{F3}) can be combined with
$F_{(2)}$ of (\ref{F2}), and we obtain for the total free energy
\beq \label{final}
F={1\over \pi \beta}\int^{(1-\delta)k_0}_0 d\omega\;
z(\beta\omega) \int^{k_0}_{\omega\over 1-\delta} dx\;
\mu({\omega\over x})H^{-1}(x).
\eeq
Note that $F$, and hence $S$, is independent of $\epsilon$, i.e.
UV finite. This result is quite general and is true as long
{\bf (i)} the transplanckian modification of the radial modes
takes the form (\ref{p-H}) with the suppression condition
(\ref{bound}), {\bf (ii)} the metric satisfies a near horizon
condition, which expressed in terms of the measure factor as
(\ref{meas-cond}). We thus see that TDR generally yields a UV
finite free energy and entanglement entropy. This is the main
result of \cite{CCL}.

  The explicit form of (\ref{final}) for the Schwarzschild metric
gives
\beq\label{xx}
F \approx  -{2k_0^3  \over 3 \pi \beta} ({2M\over \delta})^3
\int^1_0 dy \;y \ln(1-e^{-\beta k_0 y}) \tanh^{-1} y
\eeq
where we assume a large IR cutoff $2M/\delta(\gg \beta)$ to
extract the leading IR term from the $\rho$-integration. The IR
divergence is independent of the background $M$ as in the
non-transplanckian case. Moreover, the area law no longer holds
for the entropy derived from (\ref{xx}), instead, for fixed $k_0$,
both the free energy and the entropy are monotonically decreasing
functions of $\beta$ so that it can be neglected in the
semi-classical limit in comparison with the Bekenstein-Hawking
entropy.

\bigskip

\end{document}